# What Is the Temperature?
# Modern Outlook on the Concept of Temperature

Edward Bormashenko

*Chemical Engineering Department, Engineering Faculty, Ariel University, P.O.B. 3, 407000 Ariel, Israel*

ORCID id 0000-0003-1356-2486

**Abstract**

The meaning and evolution of the notion of "temperature" (which is a key concept for the condensed and gaseous matter theories) is addressed from the different points of view. The concept of temperature turns out to be much more fundamental than it is conventionally thought. In particular, the temperature may be introduced for the systems built of "small" number of particles and particles in rest. The Kelvin temperature scale may be introduced into the quantum and relativistic physics due to the fact, that the efficiency of the quantum and relativistic Carnot cycles coincides with that of the classical one. The relation of the temperature to the metrics of the configurational space describing the behavior of system built from non-interacting particles is demonstrated. The Landauer principle asserts that the temperature of the system is the only physical value defining the energy cost of isothermal erasing of the single bit of information. The role of the temperature the cosmic microwave background in the modern cosmology is discussed.

## 1. Introduction

What is the temperature? Intuitively the notions of "cold" and "hot" precede the scientific terms "heat" and "temperature". Titus Lucretius Carus noted in *De rerum natura* that "warmth" and "cold" are invisible and this makes these concepts difficult for understanding [1]. The scientific study of heat started with the invention of thermometer [2]. The operational definition of temperature is shaped as follows: temperature is what we measure with a thermometer [3]. Galileo and his contemporaries were already using thermometers around 1600. Robert Boyle, Robert Hooke and Edmond Halley suggested to use the standard "fixed points", namely phenomena that could be used as thermometric benchmarks because they are known to take place always at the same temperature. Jean-Andre de Luc and Henry Cavendish made much in order to establish these fixed points [2]. It turned out that an accurate establishment of these points poses extremely difficult experimental problems. Consider, for example the temperature, of water boiling. We have to answer exactly experimentally what is "water boiling"? Try to boil water and you will recognize that it is a complicated process, divided temporally into common boiling, hissing, bumping, explosion and bubbling. Now, which of these is true boiling? The detective story of the development of accurate thermometers is excellently reviewed in the monograph "Inventing Temperature. Measurement and Scientific Progress" [2].

The theoretical breakthrough in the understanding of temperature occurred when Carnot discovered the temperature function, which was gradually developed over a period of 30 year by Clapeyron, Helmholtz, Joule, Rankine, Thomson (Kelvin), and Clausius [4, 5]. In Thomson's final resolution of the problem, Carnot's function simply determined the "absolute" thermodynamic temperature scale [4]. Indeed, the efficiency of the Carnot engine $\eta = 1 - \frac{T_2}{T_1}$ is independent on any specific material constants and depends on the absolute temperatures of the hot $T_1$ and cold $T_2$ baths only [ 5, 6].

Let us quote William Thomson: "The relation between motive power and heat, as established by Carnot, is such that quantities of heat, and intervals of temperature, are involved as the sole elements in the expression for the amount of mechanical effect to be obtained through the agency of heat; and since we have, independently, a definite system for the measurement of quantities of heat, we are thus furnished with a measure for intervals according to which absolute differences of temperature may be estimated [2]".

And it should be emphasized that the efficiency of the Carnot engine is independent on the number of particles, constituting the working fluid of the engine [7] . This makes possible the construction of the absolute temperature scale, suggested by Kelvin, for small-scale physical systems. The non-trivial problem of matching of thermometric and absolute (Kelvin) temperature scales is discussed in detail in [6].

Remarkably, the efficiency of the Carnot cycle remains the same for a quantum mechanics cycle exploiting a single quantum mechanical particle confined to a potential well [8]. The efficiency of this engine is shown to be equal the well-known Carnot efficiency, because quantum dynamics is reversible [8]. Moreover, the efficiency of the Carnot cycle remains the same for the relativistic Carnot engine. The relativistic transformation of temperatures remain a subtle and open theme, in which different expressions for this transformation were suggested [9-10].

Planck and Einstein suggested that the transformation of temperatures is governed by: $T = T_0\sqrt{1-\frac{u^2}{c^2}} = \frac{T_0}{\gamma}$ ; $\gamma = \frac{1}{\sqrt{1-\frac{u^2}{c^2}}}$ [9]. In contrast, Ott suggested for the same transformation $T = \gamma T_0$ [10]. It was also suggested, that the universal relativistic transformation for the temperature does not exist [11-13].

However, it is easily seen that the efficiency of the Carnot cycle remains the same for the linear relativistic transformations of temperature shaped as: $T = \alpha T_0$ ; $\alpha = const$, whatever is the value of the constant. Thus, we recognize that the efficiency of the Carnot engine demonstrates remarkable stability and insensitivity to the make-up of the engine, number of the particles constituting the working fluid of the engine [8], quantum behavior of the particles and also to the motion of frameworks. This fact enables introduction of the Kelvin thermodynamic temperature scale in the realms of relativity and quantum mechanics. Somewhat surprisingly, the Carnot cycle represents the interception point for the classical, quantum and relativistic physics.

**2. Results and discussion**

**2.1. Temperature as an averaged the kinetic energy and the metrics of configurational space**

The ubiquitous understanding of temperature is that the temperature of a substance is related to the average kinetic energy of the particles of that substance [14]. It seems that this idea belongs to Daniel Bernoulli [2]. We will demonstrate that this is a very narrow definition of the temperature, and it does not always work. However, we start from this understanding of temperature and we will show that such an interpretation leads to the non-trivial relation of temperature to the metrics of the physical configurational space. Indeed, geometry enters into the realm of physics in its relation to the inertial properties of masses, in other words in its relation to their kinetic energies [15]. Consider the system of $N$ non-interacting point masses $m_i$. Their kinetic energy $E_k$ equals:

$$E_k = \frac{1}{2}\sum_{i=1}^{N}\frac{1}{2}m_i v_i^2 \qquad (1)$$

Let us define the element $ds$ of the $3N$ configurational space according to Eq. 2:

$$ds^2 = 2E_k dt^2 = dt^2 \sum_{i=1}^{N} m_i v_i^2 = \sum_{i=1}^{N} m_i(dx_i^2 + dy_i^2 + dz_i^2), \qquad (2)$$

Thus, the kinetic energy $E_k$ may be re-written as:

$$E_k = \frac{1}{2}m\left(\frac{ds}{dt}\right)^2, \qquad (3)$$

where $m_i = m = 1$. In this Euclidian configurational space $\sqrt{m}x_i; \sqrt{m}y_i; \sqrt{m}z_i$ are the Cartesian coordinates. In this space the kinetic energy of the system is represented by the kinetic energy of the single point mass with $m = 1$. When $E_k = const$ takes place this point moves with the constant velocity:

$$\frac{ds}{dt} = \sqrt{\frac{2E_k}{m}} \qquad (4)$$

Now assume that our mechanical system, built of $N$ non-interacting point masses $m_i$, is in the thermal equilibrium with a heat bath at a fixed temperature $T$ (from the statistical point of view this means that the system is described by the canonical ensemble [16]). In this case, the averaged element of the configurational space is defined as follows:

$$\overline{ds^2} = 2dt^2 \sum_{i=1}^{N} \frac{1}{2} m_i \overline{v_i^2} \qquad (5)$$

We assume that the system is ergodic and demonstrates the same statistical behavior averaged over time as over the system's entire possible state space, thus the averaging in Eq. 5 may be ever the time or ensemble averaging [16, 17]. Thus, the temperature of the system may be introduced according to Eq. 6:

$$\frac{1}{2} m_i \overline{v_i^2} = \frac{3}{2} k_B T, \qquad (6)$$

where $k_B$ is the Boltzmann constant. Hence, Eq. 5 may be re-written as:

$$\overline{ds^2} = 3N k_B T dt^2 \qquad (7)$$

and the velocity $v^*$ may be introduced:

$$v^* = \sqrt{\frac{\overline{ds^2}}{dt^2}} = \sqrt{3N k_B T}, \qquad (8)$$

(again $m = 1$ is assumed).

When the temperature of the system $T$ is constant its time evolution may be represented by the motion of the single point in the configurational $3N$-space ($\bar{x}_i = \sqrt{\overline{v_i^2}}t$; $\bar{y}_i = \sqrt{\overline{v_i^2}}t$; $\bar{z}_i = \sqrt{\overline{v_i^2}}t$) with the constant velocity $v^* = \sqrt{3N k_B T} = const$. Thus, the metrics of the configurational space is completely defined by the temperature of the system. This conclusion looks is trivial; however, actually it is not obvious due to the fact the kinetic energy of the system is not constant now, but fluctuates around the average value with the probability given by the Gibbs formula. Actually, Eq. 8 reflects the well-known input, stating that the canonical ensemble does not evolve with time [16]. Hence, the temperature of the system in the thermal equilibrium with the thermal bath defines the constant velocity of the single point (see Eq. 8), describing the motion of this point in the $3N$ configurational space, where coordinates are defined as follow: $\bar{x}_i = \sqrt{\overline{v_i^2}}\,t$ ; $\bar{y}_i = \sqrt{\overline{v_i^2}}\,t$ ; $\bar{z}_i = \sqrt{\overline{v_i^2}}\,t$.

**2.2. Temperature, energy and entropy: an alternative glance on the temperature**

An alternative looking on the temperature emerges from the concept of entropy. Alternatively, the notion "temperature" is introduced according to Eq. 9:

$$\frac{1}{T} = \left(\frac{\partial S}{\partial E}\right)_N , \qquad (9)$$

where $S$ and $E$ are the energy and entropy of the system correspondingly and $N$ is the number of particles constituting the system [3, 5, 16, 18-20]. The inverse temperature, defined according to Eq. 9 is seen as the rate of change in the entropy of the system taking place with the change in its energy. In other words the inverse temperature appears as a measure of energy necessary for ordering of the system, estimated by its entropy, which in turn is given by the Boltzmann formula $S = k_B \ln W$, where $W$ is a number of micro-states corresponding to a certain macro-state of a system [16]. Actually, this definition is very different from that supplied by Eq. 6 relating the temperature to the averaged kinetic energy of particles. First of all, it does not imply averaging and may be introduced for the system built of the arbitrary number of particles. Indeed, the notions of entropy and energy may be introduced for the physical systems containing any number of particles, whatever small or large, and even for the single-particles systems [21-22]. Secondary, it does not arise from the kinetic energy (motion) of particles. As a matter of fact, it may be successfully applied for the systems of particles in rest, such as an ensemble of spins (elementary magnets) embedded into the magnetic field [18, 20]. Thus, Eq. 9 supplies much more general definition of the temperature than that relating the concept of the temperature to the averaged kinetic energy of the system. Consider, that the Kelvin definition of the temperature, emerging from the Carnot cycle, also does not relate the temperature to the molecular motion. Substituting of Eq. 9 into Eq. 7 yields the non-trivial equation defining the averaged metrics of the configurational space:

$$\left(\frac{\partial S}{\partial E}\right)_N \overline{ds^2} = 3Nk_B dt^2 , \qquad (10)$$

and the velocity of the representing point in the configurational space:

$$v^* = \sqrt{\overline{\frac{ds^2}{dt^2}}} = \sqrt{3Nk_B \left(\frac{\partial S}{\partial E}\right)_N^{-1}} \qquad (11)$$

**2.3. The Landauer principle and informational interpretation of the temperature**

The Landauer principle establishing the physical equivalent of information supplies an additional outlook on the temperature. In its simplest meaning, the Landauer principle states that the erasure of one bit of information requires a minimum energy cost equal to $k_B T ln 2$, where $T$ is the temperature of a thermal reservoir used in the process [23-24]. Landauer also applied the suggested principle to the transmission of information and re-shaped it as follows: an amount of energy equal to $k_B T ln 2$ (where $k_B T$ is the thermal noise per unit bandwidth) is needed to transmit a bit of information, and more if quantized channels are used with photon energies $h\nu > k_B T$ [25]. Actually, the Landauer principle converted the information into physical value; Rolph Landauer himself stated that the "information is physical" [24]. The precise meaning, universality, evaluation and the interpretation of the Landauer principle were subjected to the intensive and sometimes stormy scientific discussion recently [26-36]. Whatever is the precise meaning of the Landauer Principle, it contributes to the re-construction of the fundamentals of physics on the informational basis, suggested by John Archibald Wheeler in [37] and developed in [38-41]. It immediately follows from the Landauer Principle that the temperature of the system is the *only physical value* defining the energy cost of isothermal erasing of the single bit of information [23-24, 36]. Again, recall that the temperature defined with Eq. 9 is introduced not only for the large Avogadro-number-scale systems but also for the systems built of an arbitrary number of particles [21, 22], and even for the single-particle-systems, as it is well-illustrated by the minimal Szilard Engine, whatever classical [7, 42], quantum [8] or relativistic. This means, that the mass-equivalent of the single bit of information may be introduced, which is also completely defined by the temperature of the system [27, 30, 43]. The notion of temperature, seen from the perspective of the Landauer principle and Eq. 9, is transformed into the fundamental physical quality, and it is not interpreted as "the averaged kinetic energy of particles", as it is usually understood.

**2.4. Fundamental role of the cosmic background temperature**

The fundamental role of the notion of temperature becomes even more pronounced in the context of the effect of the cosmic microwave background [44]. The discovery and interpretation of the cosmic microwave background 1965 by Arno Penzias, Robert Wilson and Robert H. Dicke was a turning point in the modern century cosmology [44]. The discovery supported the well-established now cosmological paradigm, broadly known as the Big Bang cosmology [45]. The cosmic microwave background (CMB) in the Big Bang cosmology, is electromagnetic radiation as a remnant from an early stage of the universe, also known as "relic radiation". The CMB has a thermal black body spectrum at a temperature of 2.73548±0.00057 K. It appears that this temperature today serves as one of the most important physical constants. Probably the most significant and most frequently cited consequence of the standard hot Big Bang interpretation of the CMB is the limit the background temperature sets on the fraction of universal density which can be in the form of baryonic matter. The physical picture underlying this prediction is simple: the baryonic number is (at least approximately at the timescales comparable to the Hubble time, neglecting effects of the hypothetic proton decay and other very slow processes) a conserved quantity, and the vast majority of photons currently existing in the universe are CMB photons, so the photon-to-baryon ratio today is essentially the same as it was at the time of decoupling, at redshift. Therefore, fixing the photon density per co-moving volume, coupled with limitations on the baryon-to-photon ratio in the early universe (provided by the theory of primordial nucleosynthesis; [46, 47]), gives a unique handle on the total cosmological baryon density.

### 3. Conclusions

The universal absolute temperature scale suggested by Kelvin is possible due to the amazing robustness of the Carnot cycle. The efficiency of the Carnot engine is independent on the number of particles constituting the working fluid [7] and also on the specific material constants and depends on the temperatures of the hot $T_1$ and cold $T_2$ baths only. Moreover, it remains the same for the quantum [8] and relativistic Carnot engines. This fact allows introduction of the Kelvin absolute thermodynamic temperature scale in the realms of relativity [9] and quantum mechanics [8]. This converts the notion of temperature into the key concept of the modern physics of condensed and gaseous matter. The relation of the temperature to the metrics of the configurational space describing the behavior of system built from non-interacting particles is treated. The temperature defined with the equation $\frac{1}{T} = \left(\frac{\partial S}{\partial E}\right)_N$ (where $S$ and $E$ are the entropy and energy of the system correspondingly) is not related to the averaging of the kinetic motion of particles, constituting the system, and may be introduced for the systems containing an arbitrary number of particles, including those in rest [18, 20]. From the point of view of physics of information the temperature of the system appears as the *only physical value* defining the energy cost of erasing of the single bit of information. The role of the temperature of the cosmic microwave background in constituting of basic ideas of the modern cosmology is addressed. The temperature of CMB turns out to be one of the most important fundamental physical constant limiting the value of the photon-baryon ratio in the Universe.

**Funding**

This research received no external funding.

**Acknowledgments**

The author is thankful to Yelena Bormashenko for her kind help in preparing this paper.

**Conflicts of Interest**: The author declares no conflict of interests.